# The Dynamics of Triads in Aggregated Journal-Journal Citation Relations: Specialty Developments at the Above-Journal Level

Wouter de Nooy [a]* and Loet Leydesdorff [b]

[a] Amsterdam School of Communication Research (ASCoR), University of Amsterdam, P.O. Box 15793, 1001 NG Amsterdam, The Netherlands; w.denooy@uva.nl ; *corresponding author

[b] Amsterdam School of Communication Research (ASCoR), University of Amsterdam, P.O. Box 15793, 1001 NG Amsterdam, The Netherlands; loet@leydesdorff.net

**Abstract**

Dyads of journals—related by citations—can agglomerate into specialties through the mechanism of triadic closure. Using the Journal Citation Reports 2011, 2012, and 2013, we analyze triad formation as indicators of integration (specialty growth) and disintegration (restructuring). The strongest integration is found among the large journals that report on studies in different scientific specialties, such as *PLoS ONE, Nature Communications, Nature,* and *Science.* This tendency towards large-scale integration has not yet stabilized. Using the Islands algorithm, we also distinguish 51 local maxima of integration. We zoom into the cited articles that carry the integration for: (*i*) a new development within high-energy physics and (*ii*) an emerging interface between the journals *Applied Mathematical Modeling* and the *International Journal of Advanced Manufacturing Technology.* In the first case, integration is brought about by a specific communication reaching across specialty boundaries, whereas in the second, the dyad of journals indicates an emerging interface between specialties. These results suggest that integration picks up substantive developments at the specialty level. An advantage of the bottom-up method is that no *ex ante* classification of journals is assumed in the dynamic analysis.

**Keywords**: specialty; journal; triadic closure; bottom-up; dynamic

# 1. Introduction

Scientific disciplines and specialties are difficult to delineate because their borders are diffuse and changing. Nevertheless, they are important to the study and evaluation of the sciences. For example, the citation impact of publications can be compared across subsets only after normalization, so evaluative bibliometrics requires the delineation of disciplines and specialties as reference sets for the normalization (Bornmann, Leydesdorff, & Mutz, 2013). Furthermore, the study of changing boundaries between disciplines and specialties can be expected to indicate scientific developments (Whitley, 1984). Assuming that a scientific specialty represents organization of the literature above the level of individual journals, in this study we address the question how journal-journal citation relations can be used to reveal the development of specialties within the sciences.

Clustering techniques, also known as community detection techniques, have usually been applied to networks of citations (e.g., White, Wellman, & Nazer, 2004) or to collaboration relations (e.g., Lambiotte & Panzarasa, 2009) in order to delineate scientific disciplines and specialties. Many clustering techniques have been developed both within and outside social network analysis (for an overview see, e.g., Fortunato, 2010; Newman, 2010; Wasserman & Faust, 1994). However, the different clustering algorithms and similarity criteria generate a parameter space (Ruiz-Castillo & Waltman, 2015). Different clustering solutions can thus be generated for the same network. Furthermore, clustering techniques often require parameter choices or a random seed *ex ante* that can be consequential for the results.

In the case of "big data," for example, one can search for local and global optima in this parameter space, but the results will predictably remain uncertain at the margins. The choice of one set of parameter values or another may lead to a fit (and perhaps potential validation) in some areas, but perhaps not in others (Leydesdorff & Rafols, 2011). Interdisciplinary developments can be expected to generate novelty at margins where the boundaries may then become unreliable (Rafols, Leydesdorff, O'Hare, Nightingale, & Stirling, 2012; Wagner et al., 2011). In sum, the data contain uncertainty and multivariate methods require choices of parameters.

Whereas the choice among different clustering solutions for the same network may already be difficult, the comparison of clustering solutions across years is truly a hard nut to crack (Leydesdorff & Schank, 2008; Studer & Chubin, 1980, pp. 269 ff; for a recent attempt to crack the nut, see Glänzel & Thijs, 2012). As we shall see below, for example, only a minority of citation links among journals in the Journal Citation Reports (JCR) appear in three consecutive years. As a consequence, the structure of the citation network can be expected to change considerably between subsequent years, which introduces a lot of variation in clustering outcomes even if the same clustering technique is used with the same parameter settings. Does the measurable change in comparisons between similar representations for different years indicate substantive change and development over time, or a difference in the error and uncertainty? Furthermore, clusters and journals are co-constitutive (Breiger, 1974): a cluster is identified by the journals it contains, so a change in the set of journals constituting a cluster also changes the nature of the cluster. What does it mean if a journal moves over to a different cluster (Rosvall & Bergstrom, 2010)?

Using a time-series of JCR data, we propose to abandon the focus on clustering the overall citation network and start analyzing local network structures, that is, journals in the context of the other journals with which they are directly linked. This approach resembles Cho *et al.*'s operationalization of research facilitation as increasing local network closure (Cho, Huy Hoang Nhat Do, Chandrasekaran, & Kan, 2013). Furthermore, we abandon the focus on clustering journals as nodes, and shall argue for studying the dynamics of citation links instead. The basic element of a network is a link, which is identified as a relation between a *pair* of nodes. In this study, we focus on pairs of journals (cf. Klavans, Persson, & Boyack, 2009) that are citing each other. We call this a mutual (citation) link and we restrict our analyses to this type of link.

For each pair, one can determine the density of its local network context. This can be done in several ways, but in this exploration we simply count the number of shared network neighbors of each pair. A pair of journals has a shared neighbor in the citation network if there is a third journal that they both cite and by which they are both cited. Figure 1, for example, shows the shared neighbors of the journal pair {J1, J2}: J3, J4, and J5 are mutually related to both J1 and J2. Journal J6, however, does not have a reciprocal citation link with journal J1, so it is not counted in this study as a common neighbor to journals J1 and J2. The journal pair {J1, J2} has three shared network neighbors. Each shared neighbor creates a complete or closed triad with this journal pair.

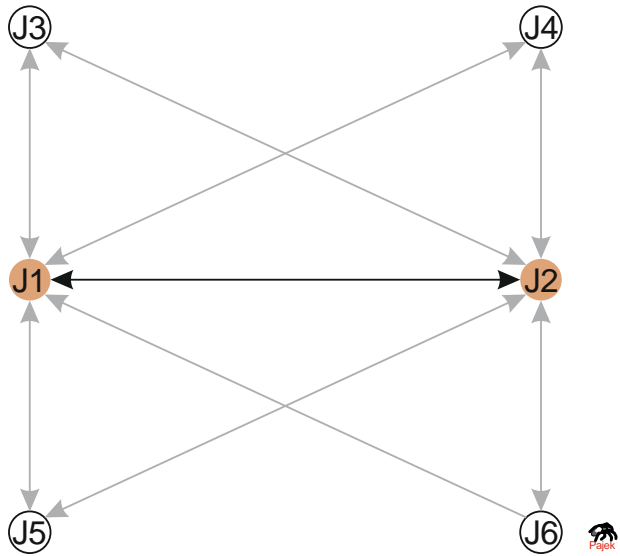
*Figure 1*. Example: Shared network neighbors of the journal pair {J1, J2}.

Already in the 1960s, social network analysts realized that two people tend to establish a direct relation more frequently if they share a network neighbor (Foster, Rapoport, & Orwant, 1963). This effect became known as the transitivity of social relations and as triadic closure, which generates a clustering of cohesive subgroups or communities at the level of the overall network (e.g., Bianconi, Darst, Iacovacci, & Fortunato, 2014; Frank & Harary, 1982). If we follow the argument that clusters at the overall network level result from local network closure, we may use triadic closure as a measure of cohesion and thus avoid the potentially problematic delineation of communities at the level of the overall network.

If a pair of journals is part of a dense cluster (cohesive subgroup) in the network, the network around the pair is also dense, which means that the two journals have more shared network neighbors. If a pair of journals has more shared network neighbors, they are also more likely to belong to a larger cluster. It is important to note that this local approach looks for centers or crystallization points of cohesion. It does not aim to establish boundaries of clusters, so it is better suited for analyzing a network with fuzzy boundaries such as those between scientific specialties.

Theoretically, the triad has been an important structural concept in social network analysis. The sociologist Simmel (1902) already argued that a social tie between a pair changes qualitatively if this pair has a shared network neighbor. Each shared neighbor creates a completely linked triad in which the pair is embedded (Krackhardt, 1999). A complete triad is more likely to be stable, to have stronger shared norms, more conformity, and less options for brokerage (Burt, 1992; Granovetter, 1973). In a similar vein, we posit that a journal pair having more shared neighbors can be expected to have a stronger and more stable focus on common themes because it is more strongly embedded in citation feedback loops through which information circulates among the journals. In our opinion, triads can be considered as the building blocks of specialty formation above the journal level.

We need triads as relatively stable building blocks because it is our aim to analyze change over time, that is, developments, and not the situation at a particular moment or period of time (e.g., Batagelj, 2012). Which pairs of journals gather more shared network neighbors, that is, a denser local network context? We associate an increase in density in terms of triads with the growth and integration of a specialty structure above the level of individual journals. In contrast, we associate a decrease in local network density with differentiation or disintegration. We do not mean to attach positive or negative connotations to these concepts because differentiation can also be a source of significant restructuring; for example, the emergence of a new specialty may cause disintegration within existing specialties.

Because one can also expect random change due to variation in the presence of citation links in different years, we focus on consistent change over time using three years of data (JCR 2011, 2012, and 2013). Which journals witness increasing local network density in each subsequent year, or decreasing local network density in each subsequent year? In this paper, *integration* is operationalized as monotonic increase over time, while *disintegration* refers to monotonic decrease. We expect these monotonic changes to reflect developments within the sciences. Ours remains a descriptive approach; it is not our aim to simulate overall network structure assuming particular rates of shared neighbor creation (Bianconi *et al.*, 2014) or to estimate this rate from longitudinal network data (Snijders, 2011), although such could be options for further research.

In some sense, we return to the focus of Leydesdorff (2004) on individual pairs of journals and local network context. Technical developments, however, now allow us to investigate the entire network of journals. In contrast to previous approaches, we can take into account all direct contacts of each journal pair in the entire set of journals without prior limitation to specific disciplines or specialties, because such an assumption would beg the question of how specialties grow, merge, or differentiate over time. How do the networks of links change, and which journal-journal relations are carrying these changes? We are able to follow up at the level of articles that carry the relations at the above-journal level.

In summary, we treat journals and their citations as a connected network without assuming a segmentation into specialties ex ante. From year to year, this network integrates around certain pairs of journals while it may become less dense elsewhere. We define the integration of a journal pair as a monotonic increase in the density of the local citation network that we operationalize by counting shared network neighbors. Each newly shared network neighbor creates a new triad. It is our objective to show that integration and disintegration, thus defined, reveal developments within and between scientific specialties. The examples that we inspect in detail suggest that local maxima rather than the overall maximum of integration are related to specialty developments.

**2. Methods and materials**

*2.1. Data*

We use three consecutive years of JCR data (2011 – 2013) of the *Science Citation Index* (SCI) because this is the most recent set of three years at the date of this research (November/December 2014).[1] We need a minimum of three time points to distinguish between monotonic increase, that is, increasing from 2011 to 2012 and from 2012 to 2013, from non-monotonic development, such as increasing from 2011 to 2012 but decreasing from 2012 to 2013. We selected all journals that were actively processed in the SCI, that is, journals for which the citations in its articles were processed in at least one of the selected three years. This yielded 8,871 unique journal names. We did not include the approximately 3,000 journals listed in the JCR of the Social Science Citation Index because one can expect a different dynamic in the social sciences.

---

[1] During the project (in November 2014), JCR 2013 data was updated by Thomson Reuters. We used the updated set. The same data are used in a companion study focusing on the generation of probabilistic entropy between the years as an indicator of change (Leydesdorff & De Nooy, in preparation).

Comparison of journal pairs over time requires journals to have unique names in all three years. We treat journals that changed their names or merged with other journals within this period as a single journal. We use the newest journal name for the entire period. In total, 90 journals changed their names or merged during this period, yielding 8,781 unique journals for our analysis. Citations to the old names were added to the new or changed names so that the database remains consistent.

Table 1. Description of the yearly citation networks (8,781 journals).

|  | 2011 | 2012 | 2013 |
| --- | --- | --- | --- |
| N of journals | 8,336 | 8,471 | 8,539 |
| N of links | 1,982,108 | 2,125,658 | 2,296,961 |
| Density (loops allowed) | .026 | .028 | .030 |
| Average degree: |  |  |  |
| - reciprocal links | 98.5 | 105.1 | 115.8 |
| - unidirectional links (no loops) | 252.7 | 272.1 | 289.7 |
| Isolates (inactive journals) | 469 | 368 | 313 |

A citation matrix was constructed for each year, containing all journals. Arcs point from the journal with the cited article, which may have been published in any year, to the journal containing the citing article, which was published in the selected year. The direction of the arcs shows the flow of information through citations. Arcs are weighted by the number of citations. The yearly citation networks are rather dense considering their size: 2.6 to 3 percent of all possible citation links are present including loops, that is, journal self-citations (Table 1). On average, a journal has reciprocal citation links with about a hundred other journals (98.5 to 115.8), and it is citing or cited by another 250 to 290 journals.

Network density and average degree increase over time, partly because fewer journals are inactive, that is, neither citing nor being cited in the JCR data. Higher density increases the number of complete triads in the network, which adds to the likelihood that the number of shared neighbors increases monotonically for a pair of journals. Because one can assume that this applies equally to all journal pairs, however, such an increase in network density is inconsequential for the detection of the pairs that indicate *relatively* high integration. We focus on these pairs.

*2.2. Methods*

The original citation network is directed, distinguishing between citing and being cited. We define the local network density of a journal pair directly connected by a reciprocal citation link as the number of journals with which both journals have a reciprocal citation link: shared network neighbors. In a network without multiple lines, this number is equal to the number of complete triads containing the journal pair. We replace reciprocal citation arcs by undirected edges in order to speed up the computation, and then calculate the number of complete triads for each reciprocal citation link using the undirected 3-rings count algorithm implemented in Pajek software for network analysis (Batagelj & Zaveršnik, 2007).[2] Using this algorithm, a new network is created for each year; each new network contains all journals, and line values express the number of shared network neighbors. Note that the original weights of the numbers of citations are disregarded in this analysis; we focus on the graph.

---

[2] The algorithm is found in Pajek 4.01 under *Network > Create New Network > with Ring Counts stored as Line Values*.

As a next step, we calculate the difference between the number of shared network neighbors for each line in two successive years, and we construct a network containing only lines for pairs of journals that show a monotonic increase or monotonic decrease in their number of shared network neighbors during both time intervals. We refer to this network as the network of monotonic change. Note that this network is undirected and weighted: The average yearly increase in the number of shared network neighbors provides the line value in this network. This line value can be positive (integration) or negative (disintegration).

In summary, a pair of journals will only be considered as part of a complete triad if both journals are citing each other. We did not calculate the number of shared network neighbors for journal pairs that do not have a reciprocal citation link. As a consequence, we can only provide a (dis)integration score for journal pairs that have reciprocal citation links in all three years. Substantively this means that we assume that the integration of a journal pair may reflect a development in science only if the two journals cite each other year after year. In our opinion, this makes sense given our research question. When a journal relation functions as the core or crystallization point of a scientific development, one can expect mutual citation.

Using the network of monotonic change, our objective is to find the journals with the strongest integration. We expect that journals with high integration scores will most vividly represent scientific developments. However, we should not focus just on the highest integration scores. The number of newly shared neighbors can be expected to increase more rapidly in dense regions because these regions contain more potentially shared neighbors. Our aim is to pay attention also to local maxima, namely journal pairs outside of the densest network region experiencing an increase in the number of shared neighbors that is larger than that of the journal pairs in their local network context. After all, new developments can also be expected at the margins of specialties.

One technique that is sensitive to relatively high integration in sparser parts of the citation network is the Islands algorithm implemented in Pajek software (de Nooy, Mrvar, & Batagelj, 2011, pp. 124-128).[3] This algorithm spots both global and local maxima among line values in a network because it identifies sets of nodes—called islands—that are linked directly or indirectly by lines that have a value exceeding the values of the lines connecting them to nodes outside the set. Metaphorically speaking, these sets of nodes rise above the nodes in their vicinity like islands above the water level. The algorithm generates a new network containing only the lines that define an island, that is, lines with the required minimum value. These lines offer a concise view of the islands. If we identify journal sets with the islands algorithm in the monotonic change network, the resulting network shows precisely those lines among the journals within a set that represent relatively many newly shared network neighbors.

Figure 2 offers an example. Journals J1 and J2 (Figure 2, top left) are linked by a line with value two whereas all of their other lines have lower line values. The two journals are connected by a stronger line among themselves than with other journals, so they form an island. Their line value is a local maximum. This applies also to journals J4 and J8 (Figure 2, top right). Journals J7, J9, and J10 (Figure 2, bottom) are directly or indirectly linked by lines with minimum value three and the values of their outside links are below three. Three is the line value that defines this island (its height) and we will only show lines with at least the defining value within an island.

---

[3] This algorithm is implemented in Pajek (4.01) under *Network > Create Partition > Islands*.

The value of the line between J7 and J9 is only one, which is less than three, so this line is irrelevant to the structure within the island. For that reason it is thin and gray instead of fat and black in Figure 2.

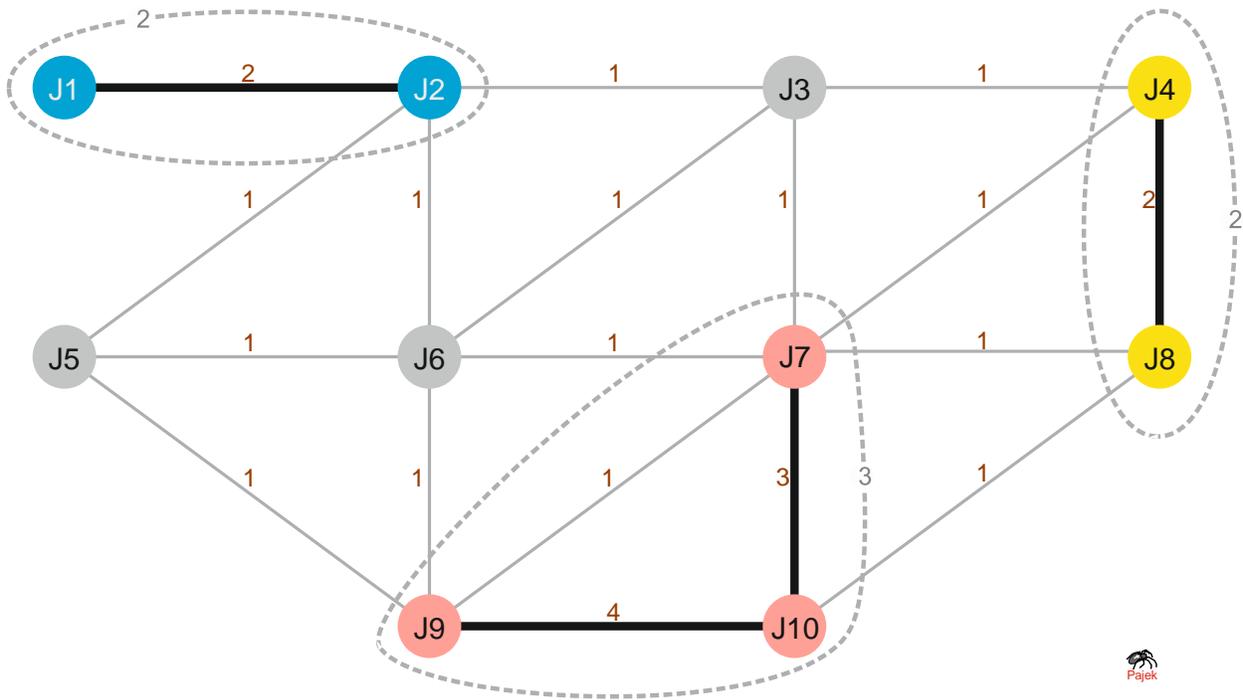

Figure 2. Results of the Islands algorithm for an example network (fat black lines define islands).

After having identified the cores of integration with the islands algorithm, we address the question of the meaning of integration in the citation network for a selection of maxima in two steps. In the first step we identify the citation links that create new complete triads. Which new citation links create newly shared neighbors? In Figure 1, for example, a new citation link pointing from journal J1 to journal J6 would make the latter journal a newly shared neighbor for the journal pair {J1, J2}. We consider a citation link to be new if it did not exist in the preceding year.

The direction and concentration of the citation links that create newly shared neighbors disclose the roles of the journals in the integration process. Do newly shared neighbors result from a distributed process in which all journals are increasingly reciprocating citations, thus completing triads, or is it a focused process in which a single journal starts citing or starts being cited by many other journals? The first type of process may reflect convergence among a set of journals, whereas the latter process signals a specific contribution by a particular journal pair to the integration.

In the second step, we identify the cited articles in the Web of Science database that are the object of the citation links identified in the first step. Which article-level citations are responsible for the newly shared neighbors? If the cited articles are very recent, it is likely that advances at a research front triggered integration. If they are not so recent, integration may signal the increasing use of another knowledge base. In the special case that the citations concentrate on a few articles, the topics of these articles may address a substantive development in science that could be responsible for integration in this citation network.

## 3. Results

If we compare citation links between journals across three years, we first note that only a minority of citation links appear every year (28 percent or 1,010,604 out of 3,569,869, see Figure 3). The number of links that appear in 2011 and 2013 but not in 2012 ($N= 235,435$) is in the same order of magnitude as the number of links that appear in two consecutive years but not in all three years (224,833 and 329,020). These results underline the instability of the yearly citation networks, which problematizes the comparison of clusters in overall network structures between years.

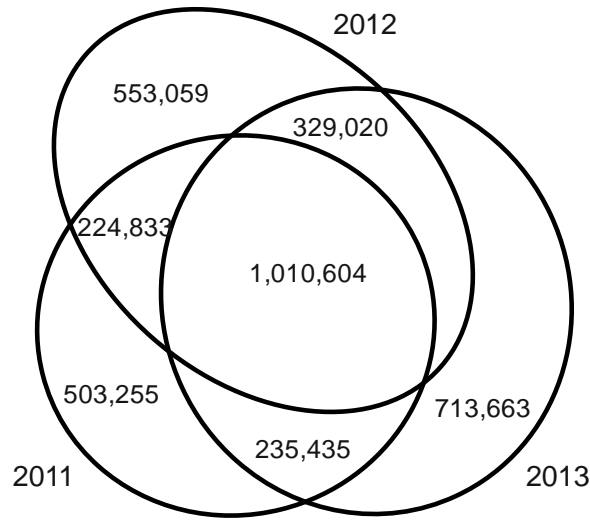

*Figure 3*. Overlap between years in number of citation links (without loops).[4]

As noted in the Methods Section, we calculate the integration of a journal pair only if the pair has a reciprocal citation link in all three years. As a consequence, the set of journal pairs that we analyze is a subset of the 1,010,604 pairs with a link in all three years (Figure 3). First, all journals pairs with a unidirectional citation link in any of the three years were omitted, leaving 244,656 journal pairs directly connected by a reciprocal citation link in all three years (Table 2). Among these journal pairs, integration is much more prevalent than disintegration: 78.8 percent of the journal pairs witness an average increase in the number of shared neighbors (Table 2). Second, all pairs without monotonic change were omitted, leaving 108,042 journal pairs with monotonic increase in the number of shared neighbors and 12,169 journal pairs with a monotonic decrease (Table 2).

---

[4] This Venn diagram was created with eulerAPE (Micallef & Rodgers, 2014).

*Table 2*. Average yearly change in shared neighbor counts for journal pairs with a reciprocal citation link in all three years, 2011-2013.

| Average yearly change in number of shared neighbors | Frequency | % |
|---|---:|---:|
| More than zero: net increase | 192,924 | 78.8 |
| *- monotonic increase* | *108,042* | |
| 0: net neutral | 5,468 | 2.2 |
| Less than 0: net decrease | 46,264 | 18.9 |
| *- monotonic decrease* | *12,169* | |
| Total | 244,656 | 100 |

*3.1. Cores of integration*

The Islands algorithm provides us with 63 cores or local maxima in the network of monotonically increasing numbers of neighbors. The largest core contains 32 journals, the next-to-largest contains six journals, and there are ten cores containing three journals and several dozens of cores of minimal size (two journals). Some of the smaller cores are defined by negative lines: they are disintegrating but at a lower rate than their neighbors in the monotonic change network if they have any neighbors at all in this network. We ignore these disintegrating cores, and then 51 cores remain in the analysis.

Some of the cores are related to one another by monotonically changing links, so that we can position the cores with respect to one another in a diagram. Figure 4 shows the largest component of these linked cores of integration. We display only the strongest integration lines within the cores, that is, the lines that define the island. Node colors indicate cores; solid lines indicate a monotonically increasing number of shared neighbors (integration); and dotted lines represent a monotonic decrease in shared neighbors (disintegration).

Two major regions of integration are indicated, in the bottom-left quadrant and the top-right quadrant. The region at the top right contains the largest core (pink – for color illustrations, see the online version of this paper), which is centered on some large and renowned multidisciplinary journals (*PLoS ONE*, *Nature Communications*, *Nature*, *Science*). It includes a smaller core (light blue) containing three journals with a focus on physical chemistry.

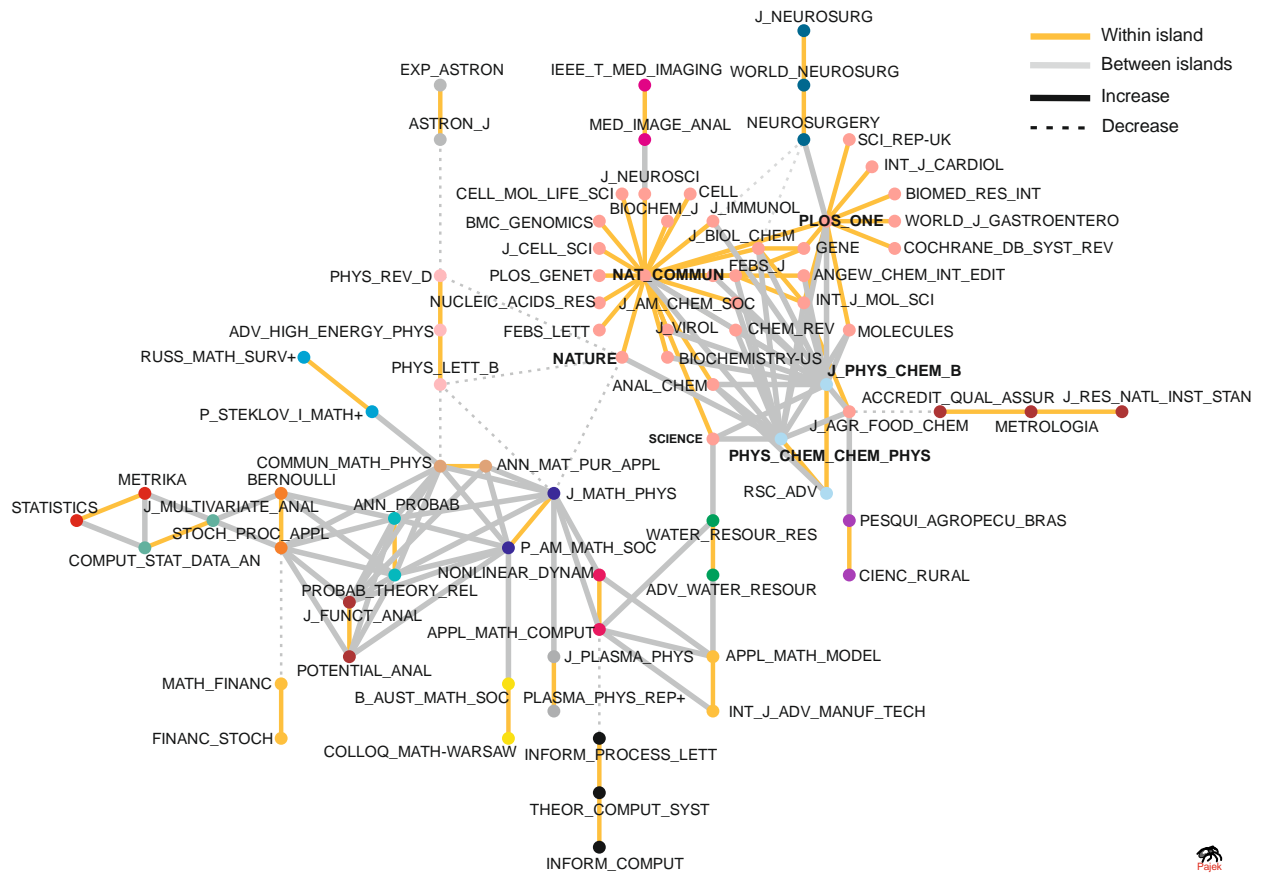

*Figure 4*. Local cores of integration and their connections, 2011-2013.

Most journals in this region of integration address genetics or biochemistry. This suggests that multidisciplinary journals increasingly cite and are cited by journals in genetics and biochemistry, or that researchers in these areas increasingly use these multidisciplinary journals as their publication outlets. Within this region of integration, there is a difference between *Nature Communications* and *PLoS ONE*, the latter journal integrating more strongly with medical journals such as the *International Journal of Cardiology* and the journal *Neurosurgery*. It is interesting to note that the journal *Neurosurgery* disintegrates from the *Journal of Immunology* and the *Journal of Biological Chemistry*, both of which integrate with *Nature*

*Communications*. This disintegration suggests that *Nature Communications* was indirectly focusing less on developments within neurosurgery during the 2011-2013 period.

The second region of integration is displayed at the bottom left of Figure 4, consisting of a series of linked local integration cores without a big core. This region mainly contains journals addressing probability theory and mathematics. It includes two journals on mathematical physics (*Communications in Mathematical Physics* and *Journal of Mathematical Physics*) which connect this region to two more isolated cores of integration: two Russian mathematical journals (*Proceedings of the Steklov Institute of Mathematics* and *Russian Mathematical Surveys*) and two journals on plasma physics (*Journal of Plasma Physics* and *Plasma Physics Reports*)

The *Journal of Mathematical Physics* is central to the relation between the two main regions of integration. On the one hand, we see that this journal and *Nature* are disintegrating, suggesting that the mathematical region is becoming less integrated with the multidisciplinary journals, perhaps as a side-effect of the latter journals' further integration with biochemistry and genetics. On the other hand, the mathematical region is indirectly linked to the multidisciplinary-biochemistry-genetics region through journals for applied mathematics or nonlinear dynamics that are integrating with journals focusing on water resources (*Advances in Water Resources* and *Water Resources Research*). The water resources journals are integrating with *Science* (but not with *Nature*). This suggests that formal models, notably in nonlinear dynamics, have led to advances in water resources research that were picked up by *Science* and by an increasing number of journals citing and being cited by both *Science* and *Water Resources Research*.

In the top-left quadrant of Figure 4, three physics journals, including a journal on high energy physics (*Advances in High Energy Physics*), are integrating, but they are disintegrating from the two large regions of integration mentioned above (the multidisciplinary/genetics/biochemistry region and the mathematics/statistics region). In addition, the physics journals are disintegrating from a local core of two astronomy journals (*Astronomical Journal* and *Experimental Astronomy*) that integrate between themselves. This suggests that these physics journals are developing in a direction that differs from the developments in mathematical physics, biochemistry and genetics, and astronomy. We will have a more detailed look at this physics core in the next section.

Before we do so, let us have a quick look at the remaining small cores, which are not directly connected by integration or disintegration links (Figure 5). In this figure, we include non-core journals that (dis)integrate with at least two core journals, so they may provide indirect links between local cores. We have labelled the non-core journals that link different cores (white nodes) in italics.

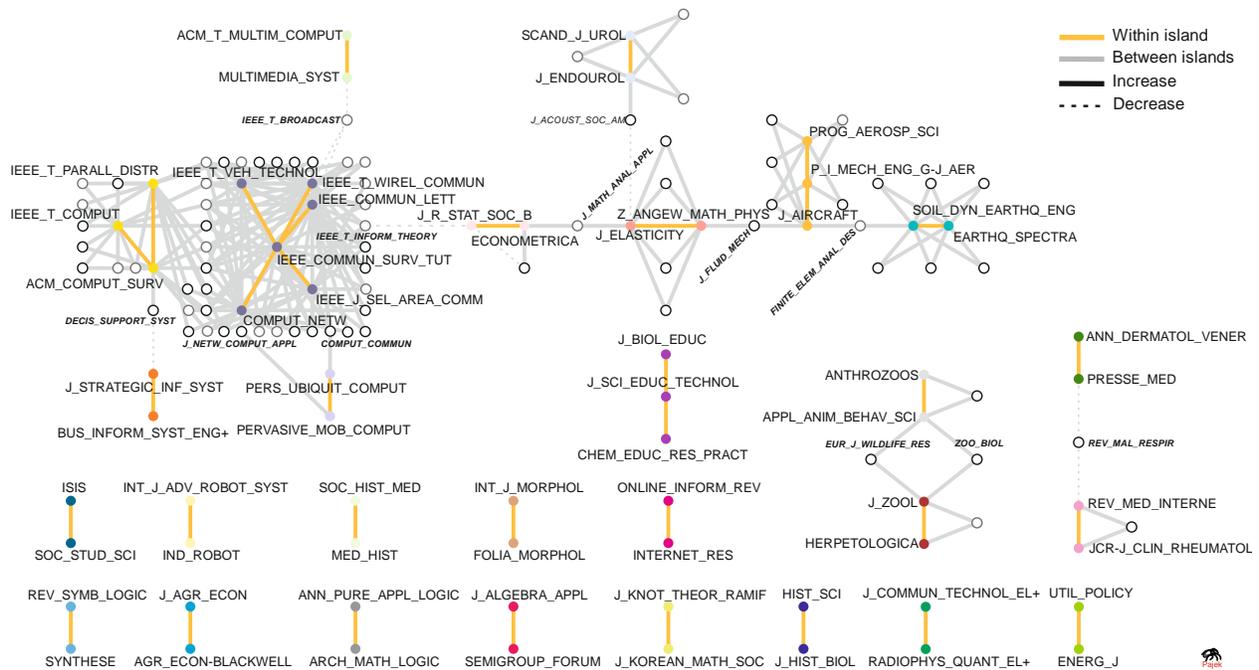

*Figure 5*. Small cores of integration that are not directly connected, 2011-2013.

The cores of integration in the top left of Figure 5, for example, feature journals in electronics and computer science with several IEEE journals as crystallization points of a core set. Two survey journals are central within the cores (*ACM Computing Surveys* and *IEEE Communications Surveys & Tutorials*), which suggests that integration reflects the aims of these journals to synthesize and sometimes reprint information from other journals within the field. These two cores are disintegrating from a string of cores (top right of Figure 5) loosely linking journals in statistics to journals in elasticity and applied mathematics/physics, on to aeronautical engineering, and finally to earthquake research. Journals in the latter two cores all belong to the same scientific specialty, as do most of the cores in the lower half of Figure 5. This result adds to the credibility of our assumption that integration measured in this way indicates developments within scientific specialties. Non-core journals that link specialized cores, for example the journal *Finite Elements in Analysis and Design* linking aeronautical engineering and earthquake research cores, may function as a source of information or methods for different specialisms.

*3.2. High energy and particle physics*

Let us take a closer look at some of the cores to assess the substantive meaning of integration. First, we focus on the three physics journals—at the top left in Figure 4—that integrate, while, at the same time, they disintegrate from other cores. *Physical Review D* (started in 1970, IF 4.86 in 2013) specializes in publications about particles, fields, gravitation, and cosmology.[5] *Physics Letters B* (started in 1966, IF 6.02 in 2013) addresses particle physics, nuclear physics and cosmology.[6] Both journals also cover high energy physics, which is the subject area of *Advances in High Energy Physics* (started in 2007, IF 2.62 in 2013).[7] First we look at the citation links that completed new triads in 2012 and 2013. Next, we look at the specific articles that are cited in these links.

In Figure 6, arcs represent citation links between journals that did not appear in the preceding year, while their appearance in the current year created a newly shared neighbor for, on the one hand, *Advances in High Energy Physics* and, on the other hand, *Physical Review D* or *Physics Letters B*. As a reading guide to this type of diagram, let us discuss this figure in more detail.

---

[5] Source: http://journals.aps.org/prd/about last accessed December 22, 2014.
[6] Source: http://www.journals.elsevier.com/physics-letters-b last accessed December 22, 2014.
[7] Sources: http://www.hindawi.com/journals/ahep/ and http://en.wikipedia.org/wiki/Advances_in_High_Energy_Physics last accessed December 22, 2014.

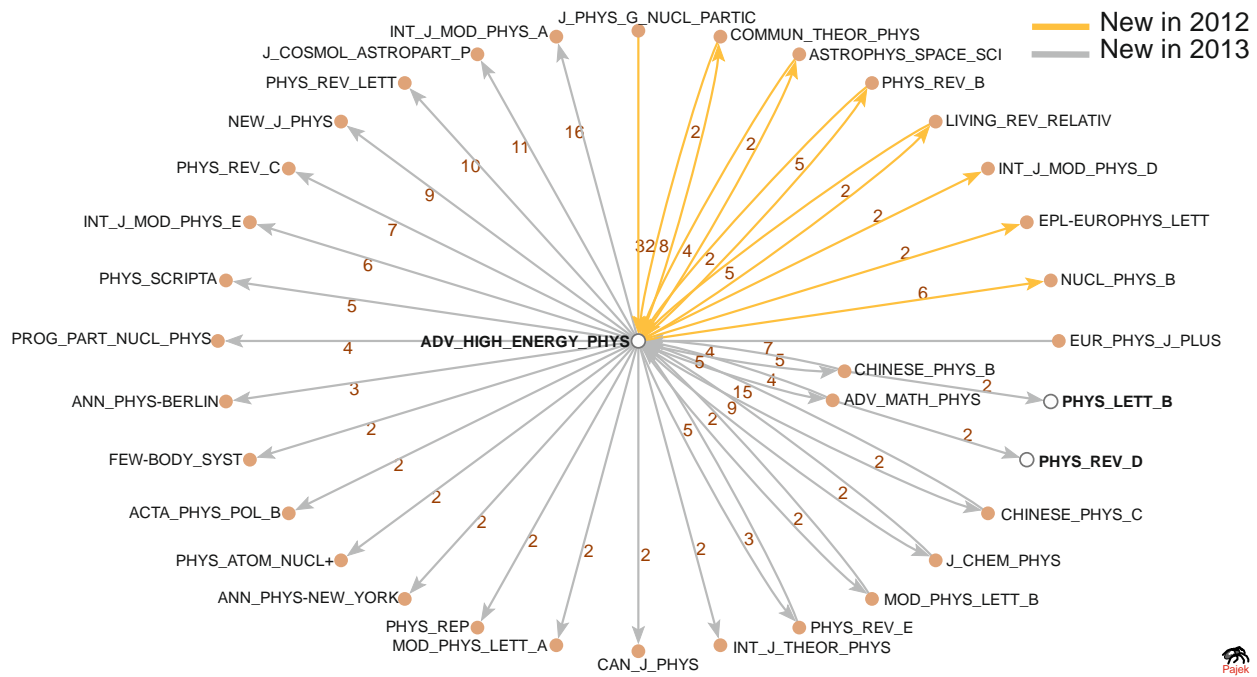

*Figure 6*. Citation relations that create new shared neighbors within the physics core (line value = number of citations).

In 2012 (orange arcs), articles in *Advances in High Energy Physics* start citing or being cited by articles in eight journals (J_PHYS_G_NUCL_PARTIC to NUCL_PHYS_B, clockwise). The new citation links create newly shared neighbors with *Physical Review D* or *Physics Letters B*, but there are no new citation links between the eight journals and *Physical Review D* or *Physics Letters B*. In other words, these eight journals already had reciprocal citation links with one or both of these journals. More generally, the absence of a citation link to or from one of the core journals in this diagram implies that this link existed both in the preceding and in the current year; otherwise the depicted arcs would not create complete triads.

For example, the unidirectional arc from the *Journal of Physics G-Nuclear and Particle Physics* to *Advances in High Energy Physics* indicates that the latter journal did not cite the former journal in 2011 but it did so in 2012. In 2012, articles in *Advances in High Energy Physics* contained 32 citations (this number is added to the arc in Figure 6) of articles published in the *Journal of Physics G-Nuclear and Particle Physics*. Absence of the reverse link indicates that the *Journal of Physics G-Nuclear and Particle Physics* must have cited *Advances in High Energy Physics* in both 2012 (otherwise there would not have been a complete triad in 2012) and 2011 (otherwise the citation link would have been new in 2012, also completing the triad). In contrast, both citations of articles in *Advances in High Energy Physics* by articles in *Communications in Theoretical Physics* and citations in the reverse direction were new in 2012, so these journals did not cite each other in 2011. Finally, *Physical Review D* and *Physics Letters B* are involved in just a single new citation link each that created a newly shared neighbor with *Advances in High Energy Physics*, namely a citation link with *Chinese Physics B* or with *Advances in Mathematical Physics* in 2013. The latter journals also established new reciprocal citation links with *Advances in High Energy Physics* in this year.

Reading Figure 6 in this way, it is clear that nearly all new network neighbors within this core of integration arose because other journals started citing articles published in *Advances in High Energy Physics*. The other two journals within this core, *Physical Review D* and *Physics Letters B*, are hardly involved in new citation links that create newly shared neighbors. Given their age, impact factor, and size—they publish thousands of articles a year versus hundreds of articles in *Advances in High Energy Physics*—these journals are likely to have stable citation links with most other journals within the subdiscipline. In contrast, *Advances in High Energy Physics* is a young and smaller journal that seems to have become more accepted as a source of information within the subdiscipline in the period 2011-2013. In the Web of Science database, it received 24 citations in 2010, increasing to 104 in in 2011, 154 in 2012, and 223 in 2013.

Citations to articles published in *Advances in High Energy Physics* have been central to the integration of these physics journals, so let us retrieve the cited articles that created newly shared neighbors. These articles appeared very recently, about 75 percent of them in the period 2011-2013. Integration thus seems to be triggered by publications at a research front (de Solla Price, 1970). One article is cited more frequently (14 times) than any other article (maximum 6 citations) by the newly shared neighbors: John McGreevy's article 'Holographic Duality with a View toward Many-Body Physics' published in 2010 (McGreevy, 2010). This article introduced researchers in condensed matter physics to a conjecture (the "anti-de Sitter/conformal field theory correspondence" conjecture) that by that time (2010) was highly cited in high energy physics.[8] This conjecture promised to make certain problems in quantum field theory mathematically more tractable.

Judging from the journals that began citing *Advances in High Energy Physics*, McGreevy's article seems to have had a much wider appeal among physicists than just among specialists in condensed matter physics. Rather than reporting on a discovery, the article makes a discovery accessible to a wider community of specialists. In this particular case, integration within the citation network thus seems to reflect growing attention to and use of a conjecture that changes the theoretical frontiers of the discipline. This may have contributed to a relatively decreased orientation on adjacent fields such as mathematics and astronomy, which would explain the disintegration around the core in this citation network.

---

[8] Source: http://www.slac.stanford.edu/spires/topcites/2010/eprints/to_hep-th_annual.shtml last accessed December 23, 2014.

*3.3. Advanced manufacturing technology*

As an illustration of quite a different situation, let us now inspect a small integration core in the nexus between the mathematical core and the multidisciplinary/biochemical/genetics core (Figure 4). The *International Journal of Advanced Manufacturing Technology* (started in 1987, IF 1.78 in 2013) aims to "bridge the gap between pure research journals and the more practical publications on factory automation systems."[9] It integrates with the journal *Applied Mathematical Modelling* (started in 1977, IF 2.16 in 2013), focusing on the mathematical modelling of engineering and environmental processes, manufacturing, and industrial systems.[10]

---

[9] Source: http://www.springer.com/engineering/production+engineering/journal/170 last accessed December 23, 2014.
[10] Source: http://www.journals.elsevier.com/applied-mathematical-modelling/ last accessed December 23, 2014.

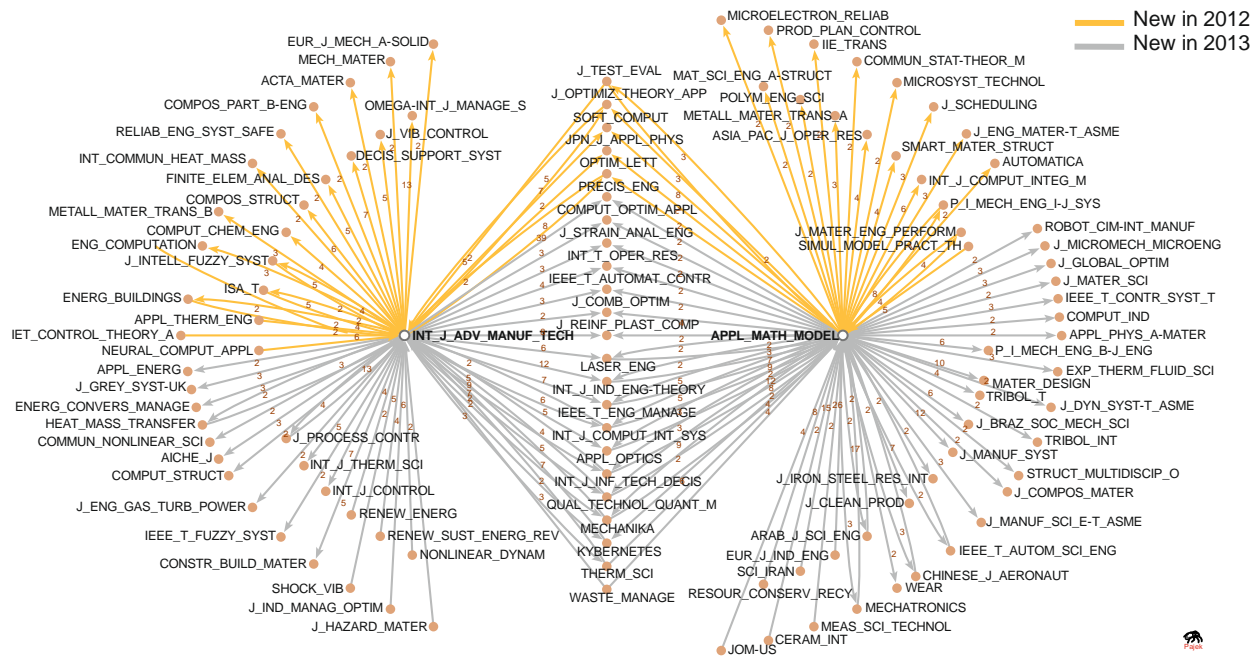

*Figure 7.* Citation relations that create new shared neighbors in the advanced manufacturing core (line value = number of citations).

In contrast to the physics cluster discussed in the preceding section, both core journals are involved in many citation links creating newly shared neighbors in 2012 and 2013 (Figure 7). As a result, we can distinguish among three sets of newly neighboring journals: journals already linked to *Applied Mathematical Modelling* but establishing reciprocal citation links with the *International Journal of Advanced Manufacturing Technology* in 2012 or 2013 (in the left of the figure), journals establishing reciprocal citation links to both core journals in the same year (in the middle of the figure), and journals already linked to the *International Journal of Advanced Manufacturing Technology* that established reciprocal citation links with *Applied Mathematical Modelling* in 2012 or 2013 (in the right of the figure). The three sets are of comparable size, so the two core journals are more or less equally involved in new citation links that boost their integration.

The arcs tend to point out from the two core journals rather than towards them. This means that new shared neighbors were created more often because the neighbors started citing articles published in the *International Journal of Advanced Manufacturing Technology* (62 new shared neighbors) or *Applied Mathematical Modelling* (51 newly shared neighbors) and not the other way around (25 and 24 newly shared neighbors, respectively). Integration accelerated in 2013 because a majority of newly shared neighbors were established in 2013 (grey arcs), especially among the journals that did not previously have a reciprocal citation link with either of the two focus journals (the group in the middle of Figure 7). Especially among the latter group of journals, the new citation links are quite often reciprocal, which indicates that the core journals and the newly shared neighboring journals started citing each other in the same year.

These results suggest that articles published in the *International Journal of Advanced Manufacturing Technology* or *Applied Mathematical Modelling* have been central to the integration of journals in this region of the citation network. Metaphorically speaking, the dyad of two journals has acted as a crystallization point for integration. In this case, however, newly shared neighbors were not created by particular articles. No single article published in *Applied Mathematical Modelling* was cited more than four times, and no single article in the *International Journal of Advanced Manufacturing Technology* was cited more than twice in the context of citation links that created newly shared neighbors. The cited articles do not hint at a particular discovery or path-breaking article that could be responsible for the integration.

Furthermore, only 30 percent of the cited articles were published in the period 2011-2013, while about 65 percent were published in the first decade of the 21th century; the oldest cited article was published in 1981. In comparison to the high energy/particle physics example, integration between the *International Journal of Advanced Manufacturing Technology* and *Applied Mathematical Modelling* seems to result more from a backlog of articles than from recent developments at a research front. An increasing number of journals are using the knowledge base stored in these journals.

## 4. Conclusions

Citations among articles, authors, and journals reflect influences and substantive similarities, so they offer great material for studying the structure and development of the sciences. Citations establish links, so they can be fruitfully studied as networks. Current computational power and routines for network analysis make it feasible to analyze the structures of networks within a chosen time interval. Cross-sectional descriptions of the overall structure of citation networks, however, do not lend themselves well to detecting longitudinal developments.

In this study, we have taken a local (bottom-up) approach to clustering by analyzing single citation links within their local citation network context. A citation link represents two journals that cite each other's articles. We defined the integration of a citation link as a monotonic increase in the number of network neighbors that the two journals share over a three-year period. The number of shared network neighbors is a measure of triadic closure. We defined disintegration as a monotonic decrease in the number of shared network neighbors over the three-year period. We focused on local maxima of integration, viz., pairs of journals and sets of journals that integrate more strongly between themselves than with other journals.

We expect that integration is related to developments in the sciences and that local maxima of integration enable us to deconstruct the nature of these developments. Integration within the citation network as we defined it seems to correspond with the substantive integration of scientific knowledge in various respects. Well-known multidisciplinary journals that aim to survey recent developments within science at large, such as *Nature*, *Science*, *Nature Communications*, and *PLoS ONE*, are indeed pivotal to the strongest integration in the citation network. These journals are likely to integrate different scientific disciplines and specializations, so this result did not come as a surprise.

The added value of our integration measure is that we can now see that genetics, biochemistry, and biophysics are the subdisciplines that have been integrating further in the 2011-2013 period. In addition, our measure shows that the integration of the large multidisciplinary journals is still increasing and has not yet stabilized. This result suggests a hypothesis for further research, namely that integration results here from a new journal business model with a different type of peer review system. If this hypothesis is correct, our measure of integration also picks up a type of integration that is not related to specialties at the journal level or to the papers' contents more specifically. This type of integration involves the largest journals, so it is likely to produce the highest integration scores. Substantive developments should then be looked for among lower integration scores, which underlines the importance of taking into account local optima.

An example of disciplinary integration was found among journals in high energy and particle physics. In this example, we could pinpoint a particular journal (*Advances in High Energy Physics*) and a specific article published in this journal as the source of many newly shared neighbors among a set of physics journals. According to its abstract, the referenced article was "[…] based on a series of lectures […]. The goal of the lectures was to introduce condensed matter physicists to the AdS/CFT correspondence." (McGreevy, 2010). The paper tried to make some very complex and specialized information accessible to other researchers, who increasingly used it among their references. Here, substantive integration should be understood as a widening of the circle of researchers who use specific information from a theoretical research front.

A different type of substantive local integration would apply to the second case that we investigated in detail, the local integration between applied mathematical modelling and advanced manufacturing technology. Here, the back catalogue of articles published by the two journals is increasingly used by other journals as part of their knowledge base. Due to their age, the articles are less likely to represent developments at a research front, and there were no specific articles that created many newly shared neighbors. Information from the interface between these journals seems to spread among journals focusing on neighboring substantive areas.

## 5. Discussion

It goes without saying that an instrument for picking up specialty development in the sciences can be of great value to scientists and organizations developing science policies. We have presented some results indicating that our integration measure picks up developments above the level of individual journals. Our instrument, however, need not detect all developments or even all salient developments. It would be informative to trace some recent scientific developments that can be considered as important from the perspective of hindsight, and test whether this integration measure picks them up and, if so, at what stage in their development. In such a design, the practical consequences of the technical choices that we have made can also be evaluated.

For example, the choice to calculate the integration of a link, that is, a pair of journals instead of a single journal, excludes the possibility of finding a substantive development that is still confined to one journal. Is this a problem, or do scientific developments quickly spread beyond the borders of a single journal? The condition that the two journals for which the integration score is calculated must have a reciprocal citation link may conceal new developments within less dense regions of the citation network. But relaxing this condition would also produce an overwhelming number of integrating pairs that are probably not related to substantive developments. Finally, case studies may show us whether it is sufficient to monitor monotonic increase over three subsequent years; developments may be much faster or slower.

Our results give us some confidence that integration measured in the proposed way offers a useful heuristic for tracing substantive developments within the sciences at the specialty level. Our approach addresses a communicative dynamic that abstracts from the information carriers; it does not matter which journal published a paper, or which paper contained which citations or was cited. What matters is that information became available and was shared at the network level. This is in line with our model of science as a communication system.

Although we address a communicative dynamic that abstracts from the information carriers, our approach enables us to zoom in to the information that was shared because we can see which journals, papers, and citations were carrying the communicative dynamic. Thus we can decompose the network relations in terms of events and provide the changes in the system with a substantive interpretation. More cases could be explored and longer periods could be examined to strengthen our confidence in using this method for monitoring the dynamics of the sciences.

**Acknowledgement**

We thank Thomson-Reuters for providing us with the JCR data.